\documentclass[aps,prl,reprint,amsmath,amssymb,showpacs,superscriptaddress]{revtex4-2}
\usepackage[utf8]{inputenc}
\usepackage[T1]{fontenc}

\usepackage{bm}
\usepackage{graphicx}
\usepackage{float}
\usepackage{color}
\usepackage{multirow}
\usepackage{xcolor}
\usepackage{booktabs}

\usepackage[colorlinks,allcolors=blue]{hyperref}
\usepackage{orcidlink}
\makeatletter
\def\NAT@def@citea{\def\@citea{\NAT@separator}}
\makeatother

\begin{document}
%\begin{bibunit}
    
\title{{\it Ab initio}  charge form factors and radii of light isoscalar nuclei:\\ Role of the two-body charge density}

\author{Xiang-Xiang Sun\orcidlink{0000-0003-2809-4638}}
\email{x.sun@fz-juelich.de}
\affiliation{Institute for Advanced Simulation (IAS-4), Forschungszentrum J\"{u}lich, D-52425 J\"{u}lich, Germany}

\author{Vadim Baru\orcidlink{0000-0001-6472-1008}}
\email{vadimb@tp2.rub.de}
\affiliation{Institut f\"ur Theoretische Physik II, Ruhr-Universit\"at Bochum, D-44780 Bochum, Germany}

\author{Arseniy A. Filin\orcidlink{0000-0002-7603-451X}}
\email{arseniy.filin@ruhr-uni-bochum.de}
\affiliation{Institut f\"ur Theoretische Physik II, Ruhr-Universit\"at Bochum, D-44780 Bochum, Germany}

\author{Evgeny Epelbaum\orcidlink{0000-0002-7613-0210}}
\email{evgeny.epelbaum@ruhr-uni-bochum.de}
\affiliation{Institut f\"ur Theoretische Physik II, Ruhr-Universit\"at Bochum, D-44780 Bochum, Germany}

\author{Hermann Krebs\orcidlink{0000-0002-2263-0308}}
\email{hermann.krebs@rub.de}
\affiliation{Institut f\"ur Theoretische Physik II, Ruhr-Universit\"at Bochum, D-44780 Bochum, Germany}

\author{Ulf-G. Mei{\ss}ner\orcidlink{0000-0003-1254-442X}}
\email{meissner@hiskp.uni-bonn.de}
\affiliation{Helmholtz-Institut~f\"{u}r~Strahlen-~und~Kernphysik~and~Bethe~Center~for~Theoretical~Physics, Universit\"{a}t~Bonn,~D-53115~Bonn,~Germany} 
\affiliation{Institute for Advanced Simulation (IAS-4), Forschungszentrum J\"{u}lich, D-52425 J\"{u}lich, Germany}
\affiliation{Peng Huanwu Collaborative Center for Research and Education, International Institute for Interdisciplinary and Frontiers, Beihang University, Beijing 100191, China}
\affiliation{CASA, Forschungszentrum J\"{u}lich, 52425 Ju\"{u}lich, Germany}

\author{Andreas Nogga\orcidlink{0000-0003-2156-748X}}
\email{a.nogga@fz-juelich.de}
\affiliation{Institute for Advanced Simulation (IAS-4), Forschungszentrum J\"{u}lich, D-52425 J\"{u}lich, Germany}
\affiliation{CASA, Forschungszentrum J\"{u}lich, 52425 Ju\"{u}lich, Germany}

\date{\today}
\begin{abstract}
We make \textit{ab initio} predictions of charge form factors (FFs) and radii for the isoscalar nuclei  $^6$Li and $^8$Be using the Jacobi-coordinate No-Core Shell Model.
The calculations employ
chiral semilocal momentum-space regularized two- and three-nucleon interactions, together with consistently regularized one- and two-nucleon electromagnetic charge operators. 
With the short-range charge density fixed to the $^4$He charge radius, the predicted FFs and the $^6$Li radius show good agreement with available experimental data. 
We find that two-nucleon charge density contributions are essential for describing the FFs, particularly at intermediate and large momentum transfers.
Although their influence on the charge radii is limited, these contributions remain crucial for attaining accurate
predictions.
The present results highlight the importance of two-nucleon charge operators in addressing the long-standing underestimation of nuclear charge radii in \textit{ab initio} calculations based on modern chiral interactions.
\end{abstract}
\maketitle

%*********************************************************%
%---------------------Introduction------------------------%
%*********************************************************%
{\it Introduction}---Electromagnetic probes offer a uniquely incisive
tool for investigating nuclear structure 
\cite{Donnelly:1984rg,Gilman:2001yh,Bacca:2014tla,Marcucci:2015rca}. 
The weak nature of the electromagnetic coupling (characterized by the fine-structure constant) simplifies the reaction mechanism, establishing a clear and direct link between experimentally measured cross sections and computed nuclear structure properties 
\cite{Donnelly:1984rg,Krebs:2020pii}. 
Elastic electron scattering experiments, in particular, yield crucial information on a nucleus's charge and magnetic form factors (FFs), which are functions of the momentum transfer squared ($Q^2$), and reflect the spatial distributions of charge and magnetization within the nucleus. 
The charge radius, derived from the charge form factor at zero momentum transfer, serves as a fundamental measure of the spatial extent of the nuclear charge distribution. 
Recent breakthroughs in experimental techniques, 
including ultra-high precision atomic spectroscopy measurements, 
have provided charge radii and magnetic moments with unprecedented accuracy
\cite{CREMA:2025zpo,vanderWerf:2025ovr,Krauth:2021foz,CREMA:2016idx}. 
These high-precision data from various advanced experimental techniques provide stringent tests for theoretical models of nuclear forces and currents, as well as for nuclear structure %especially for charge radii
\cite{GarciaRuiz:2016ohj,Gorges:2019wzy,Koszorus:2020mgn,Yang:2022wbl,Konig:2023rwe,Warbinek:2024ncq,Bai:2025dbi,Gustafsson:2025xpd}.
The quest for a profound understanding of the nuclear structure and dynamics of atomic nuclei stands as a central, 
long-standing challenge in nuclear physics 
\cite{Hergert:2016etg,Bissell:2016vgn,Lapoux:2016exf,GarciaRuiz:2016ohj,Koszorus:2020mgn,Kaur:2022yoh,Maris:2020qne,LENPIC:2022cyu,Maris:2023esu,Elhatisari:2022zrb,Ren:2025vpe}. 

Atomic nuclei are complex many-body systems composed of protons and neutrons, 
bound together by the strong nucleon-nucleon interaction. 
Chiral Effective Field Theory ($\chi$EFT) rooted in QCD serves as a robust framework for %systematically 
describing nuclear forces and currents
\cite{Epelbaum:2008ga,Machleidt:2011zz,Barrett:2013nh,
Hebeler:2015hla,Hergert:2015awm,Phillips:2016mov,Krebs:2020pii}. 
This approach enables systematic improvements in theoretical precision and allows for quantifying uncertainties stemming from neglected higher-order contributions. 
$\chi$EFT is crucial to methodically 
explore the internal structure and dynamics of atomic nuclei through electromagnetic observables.
A key advantage of $\chi$EFT lies in its ability to provide electromagnetic current operators that are consistent with nuclear forces 
\cite{Gilman:2001yh, Marcucci:2015rca, Kolling:2009iq, Kolling:2011mt, Krebs:2019aka, Krebs:2020pii, Pastore:2008ui, Pastore:2009is, Pastore:2011ip}. 
To achieve reliable predictions, both nuclear potentials and current operators must be derived and regularized within the same framework,
while inconsistent regularization, for example, can lead to violations of chiral symmetry \cite{Krebs:2020pii}. 
The use of consistent regularization is essential for obtaining precise results and for validating the theoretical framework by, e.g., demonstrating that observables are independent of unphysical effects like unitary ambiguities 
\cite{Krebs:2020pii, Filin:2020tcs}. 
Recent developments in $\chi$EFT have focused on implementing consistent regularization methods for both nuclear forces and currents. In particular, a rigorous regularization scheme in $\chi$EFT, which preserves chiral and gauge symmetries, has been developed in \cite{Krebs:2023ljo, Krebs:2023gge} using the chiral gradient flow. This approach is currently being applied to derive the subleading three-nucleon forces and loop contributions to the nuclear charge and current operators. The isoscalar contributions to the two-body charge density needed in this study, however, involve only tree-level contributions, whose regularized expressions can be found in \cite{Filin:2020tcs}.

Precise calculations of charge FFs and radii have been achieved for the deuteron using advanced high-order nuclear interactions from $\chi$EFT with consistently regularized densities and current operators, demonstrating that two-body charge operators play an important role \cite{Filin:2019eoe, Filin:2020tcs}.
Preliminary results from a combined analysis of light nuclei with $A\le 4$ suggest similar conclusions \cite{Filin:2025Marciana}. 
However, the role of two-body operators in the charge FFs and  radii of heavier
nuclei are still unclear.
As an important step towards the application to complex nuclei, this work aims to determine
the charge FFs and radii for $p$-shell nuclei $^6$Li and $^8$Be, including two-nucleon currents and reveal their importance with increasing system size. 

{\it Theoretical framework}---The nuclear many-body wave functions are obtained by using the Jacobi-NCSM 
\cite{Liebig:2015kwa, Le:2020zdu, Le:2022ikc, Le:2021gxa, Le:2023bfj}
with the SRG evolved semilocal momentum-space-regularized (SMS)
NN potential at the order N$^{4}$LO$^{+}$ 
with momentum cutoffs $\Lambda_N=$ $450$ and $500$~MeV \cite{Reinert:2017usi}, 
together with a consistent 3N potential at N$^{2}$LO 
(SMS N$^{4}$LO$^{+}$ + N$^{2}$LO), which provide
a nearly perfect description of the binding energies of 
light nuclei \cite{Maris:2020qne, LENPIC:2022cyu}. 
Along with these SMS interactions, 
the consistently regularized charge density has been developed for the isoscalar channel 
\cite{Krebs:2020pii,Filin:2019eoe,Filin:2020tcs}
and the involved low-energy coupling constants (LECs)
can be determined from the light-nuclei data
as described below.  
%(see Supplemental Material  \cite{supp} for details).

The charge FFs of the ground state are obtained by
\begin{equation}
\begin{split}
&\langle J M' |\hat {O}(\vec q, \vec k) | J M \rangle \\
& 
=
\langle JM' | 
\begin{pmatrix}
A \\ 1
\end{pmatrix}
{\hat O_\mathrm{1b}}(\vec q, \vec k) +
\begin{pmatrix}
A \\ 2
\end{pmatrix}
{\hat O_\mathrm{2b}}(\vec q, \vec k) | JM \rangle ,
\end{split}
\label{eq:ffs}
\end{equation}
where $\vec k$ is the incident momentum of the probe, $\vec q$ is the momentum transfer, $J$ is the spin of the ground state and $M'$ and $M$ are the projections of $J$. 
The charge FFs of $A \ge 4$ isoscalar nuclei are calculated using 
the transition-density formalism \cite{Griesshammer:2020ufp}.
In this way, 
the matrix of one-body ($\hat{O}_\mathrm{1b}$) and two-body ($\hat{O}_\mathrm{2b}$) charge density operators in different partial waves
calculated using a numerical partial wave decomposition based on 
Ref.~\cite{numpwd:2022},
are convoluted with one- and two-body transition densities 
encoding the nuclear many-body wave functions.
The densities are
constructed based on the J-NCSM wave function and available online at 
\cite{densdb:2025nuc}. More details can be found in Sec.~\ref{sec:method}
in~\cite{supp}.
The nucleon FFs in the charge density operators are taken from the dispersive analysis of 
Ref.~\cite{Lin:2021xrc}, which yields a proton charge radius consistent with the CODATA-2022 value.

The charge radius for an isoscalar nucleus
is composed of the contribution of the 
matter radius,
the proton and neutron charge radii, Darwin-Foldy (DF), spin-orbit (SO), two-body contact, and the one-pion-exchange ($1\pi$) terms
\begin{equation}
R_c^2 = r_m^2 + r_p^2 +r_n^2 + r_\mathrm{DF}^2 + r_\mathrm{SO}^2 + r_{1\pi}^2 + r_\mathrm{Cont}^2,
\label{eq:rc}
\end{equation}
where the proton charge radius is $r_p =0.84075(64)$ fm
from the CODATA-2022 \cite{Mohr:2024kco} and the mean-squared neutron charge radius $r_n^2 = -0.105^{+0.005}_{-0.006}$ fm$^2$ is taken from Refs.~\cite{Filin:2019eoe,Filin:2020tcs}.

{\it Determination of the LECs---}
For the two-body charge densities, the $1\pi$ term has no free parameters, while the isoscalar short-range part depends on
three LECs. 
The deuteron charge and quadrupole FFs
were used in \cite{Filin:2019eoe, Filin:2020tcs} to determine the two LECs in the isospin-zero-to-isospin-zero channel for the SMS interactions we employ here. 
In this work, we fix the remaining LEC in the 
isospin-one-to-isospin-one channel by fitting it to the experimental value of the  $^4$He
charge radius, see Sec.~\ref{sec:S1} in the Supplemental Material \cite{supp} for details.

\begin{figure}[!htbp]
    \centering
    \includegraphics[width=\linewidth]{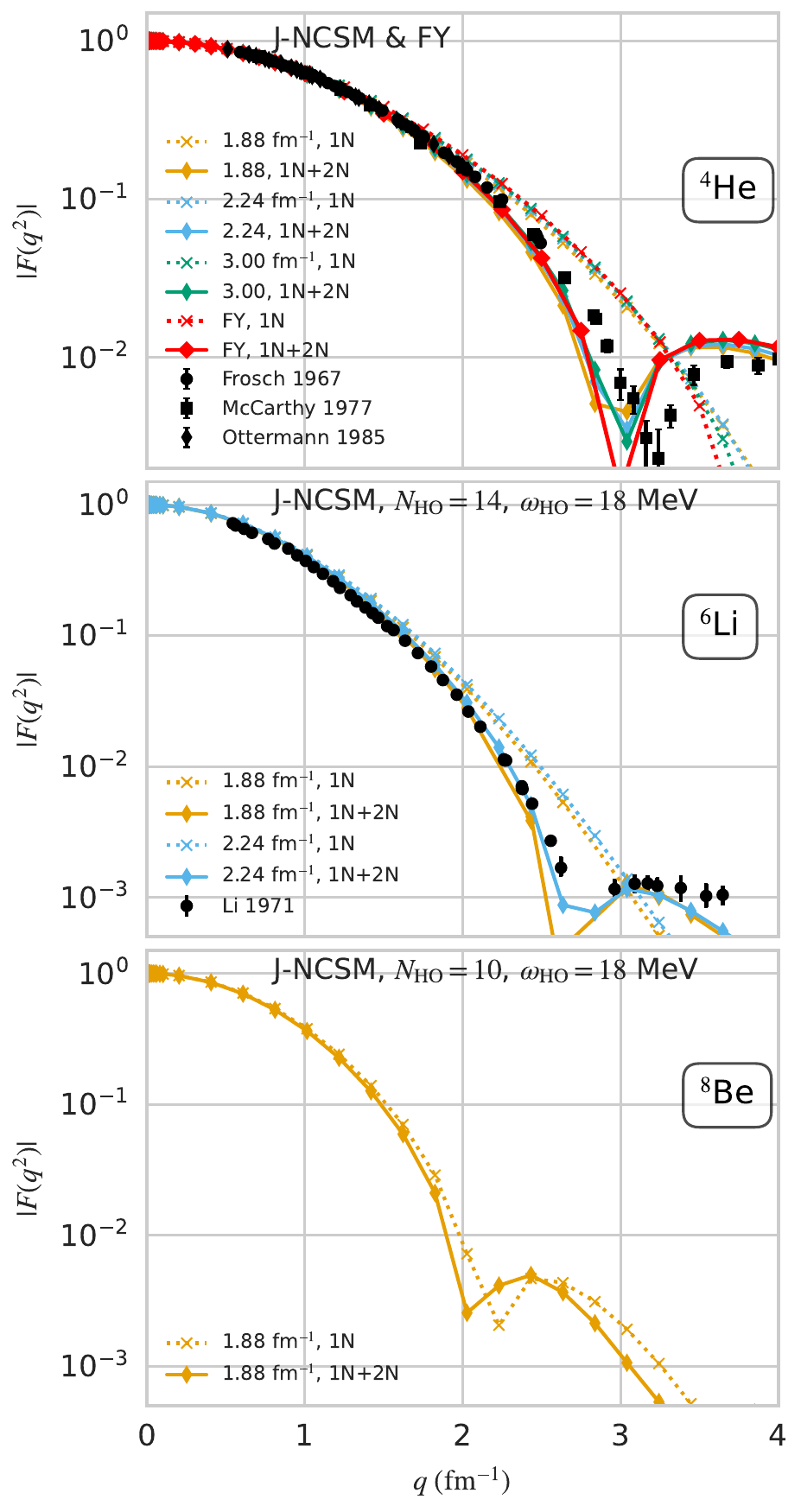}
    \caption{Charge form factors for $^{4}$He, $^6$Li and $^8$Be with SMS interaction $\Lambda_N=450$ MeV. The experimental data of $^{4}$He
    are taken from Frosch~1967 \cite{Frosch:1967pz}, 
    McCarthy 1977 \cite{Mccarthy:1977vd},
    Ottermann 1985 \cite{Ottermann:1985km}.
    Data for $^6$Li is from 
    Li 1971 \cite{Li:1971tk}. 
    }
    \label{fig:ffs}
\end{figure}

{\it Charge form factors and radii}---
The charge FFs for $^4$He, $^6$Li and $^8$Be, calculated using the SMS interaction with $\Lambda_N=450$~MeV and taking into account the one-body (1N) and the combined one- and two-body (1N+2N) contributions to the charge FFs are shown in Fig.~\ref{fig:ffs}. 
In the J-NCSM calculations, we use the SRG evolved SMS interactions and the 
corresponding one-body density matrix. 
For the two-body densities, we apply a unitary transformation to the J-NCSM wave function to eliminate the influence from the SRG \cite{Sun:2025aci}.
The inclusion of two-body currents significantly affects the form factor in the intermediate- and large-$q$ regions, shifting the  diffraction minimum to lower values of $q$,
thereby improving the agreement with experimental data for $^4$He and $^6$Li. 

For the benchmark nucleus $^4$He, J-NCSM calculations yield converged 
results with a large model space, which can be compared with charge FFs 
using the density matrix from the Faddeev-Yakubovsky (FY) method with
the bare SMS interactions.
From Fig.~\ref{fig:ffs}, it is clear that
with the increase of the SRG flow parameter $\lambda$ \cite{lambda} from $1.88$~fm$^{-1}$ to $3.00$~fm$^{-1}$, the results get closer to the FY calculations.  
The tiny differences mainly originate from the neglected SRG transformation on the one-body density (see the deviations of the 1N part above 3.5~fm$^{-1}$) and the center-of-mass motion between the two-body pair and the remaining cluster. In result, the FFs from J-NCSM
are slightly smaller than those from FY when momentum transfers are smaller than the diffraction minimum (around 3.2~$\mathrm{fm}^{-1}$).

For $^{6}$Li, we calculate its FFs with different basis properties:
the harmonic oscillator (HO) frequency $\omega$ and the maximal HO excitation $N_\text{HO}$ (see Supplemental Material \cite{supp}).
The dependence on the model space parameters turns out to be insignificant. Therefore, in Fig.~\ref{fig:ffs}, we only show the results using
$\omega_\mathrm{HO}=18$ MeV and $N_\text{HO}=14$ with two SRG flow parameters, for which the binding energy of $^6$Li is well reproduced. 
The calculated FFs
are consistent with the data when the 2N contributions are included, with only a slight underestimation observed in the region $2<q<3$ fm$^{-1}$. To further investigate the relative importance 
of the 2N charge density contributions, we also consider the unbound $^8$Be nucleus. In this case, 
direct comparison to experimental data is not possible, as no data are available. However, comparing the 1N and 2N charge density contributions indicates that 
their relative size stabilizes as the system size increases. 
Additionally, we also performed calculations with the SMS interaction $\Lambda_N=500$~MeV and the same conclusions are obtained.

\begin{table*}[tbp]
    \caption{\label{tab:rc}Charge radii of isoscalar nuclei $^4$He, $^6$Li, and $^8$Be using SMS interactions with $\Lambda_N=$ 450 MeV and 500 MeV. The contributions from the matter radius $r_m$, the DF and SO terms $\Delta R_\mathrm{DF+SO}$, and two-body contact and one-pion-exchange operators $\Delta R_{\mathrm{Cont}+1\pi}$, are shown. $r_m^\mathrm{SRG}$ labels
    the matter radius calculated by using SRG evolved two-body densities at the zero-momentum transfer and $R_c$ is the corresponding charge radius. $\tilde{R}_c$ denotes the charge radius including the tail correction of the SRG-evolved two-body relative density distribution. Available experimental data (Exp.)~are also shown. 
   The last column shows the extrapolated energies using the exponential function for the J-NCSM calculations in comparison with data from Ref.~\cite{Wang:2021xhn} and the FY calculations. 
    }
    \begin{ruledtabular}
    \begin{tabular}{llllllllll}
              & $\lambda$ (fm$^{-1}$)        & $r_m$ (fm)  & $r^\mathrm{SRG}_m$ (fm)  &  $\Delta R_\mathrm{DF+SO}$ (fm) & $\Delta R_{\mathrm{Cont}+1\pi}$ (fm) & $R_c$ (fm)& $\tilde R_c$ (fm) & $E_b$ (MeV) \\ \hline
    $^4$He  &\multicolumn{7}{l}{N$^4$LO${^+}$+N$^2$LO ($\Lambda_N=450$ MeV)}  \\
             & 1.880    & 1.4665  &1.4467 &  0.0097&0.0432&1.6945& 1.6945& 28.32  \\
             & 2.236    & 1.4524  &1.4419 &  0.0097&0.0428&1.6899& 1.6899& 28.24\\
             & 3.000    & 1.4411  &1.4372 &  0.0097&0.0421&1.6850& 1.6850& 28.22 \\  
             & bare, FY     & 1.4315 & & 0.0096&0.0404& &$1.6782$\footnote{\label{notefit}Value fitted to the experimental data for the $^4$He charge radius.}& 28.32 \\ 
     &\multicolumn{7}{l}{N$^4$LO${^+}$+N$^2$LO ($\Lambda_N=500$ MeV)}  \\
             & 1.880   & 1.4691 &1.4510 &0.0098 &0.0457 & 1.7008&1.7008& 28.34\\
             & 2.236   & 1.4559 &1.4467 &0.0098 &0.0454 & 1.6967&1.6967& 28.23\\
             & 3.000   & 1.4429 &1.4380 &0.0097 &0.0450 & 1.6886&1.6886& 28.23\\  
             & bare, FY&1.4295        &&0.0095      & 0.0422      &      &    $1.6782$\footref{notefit} &  28.53\\
             & Exp.    &        &       &       &      & & 1.67824(83) \cite{Krauth:2021foz}  & 28.29 \cite{Wang:2021xhn}  \\  \hline  
    $^{6}$Li &\multicolumn{7}{l}{N$^4$LO${^+}$+N$^2$LO ($\Lambda_N=450$ MeV), $\omega_\text{HO}=18$ MeV, $N_\text{HO}=14$ }   \\
             & 1.880  &2.2094  &2.1990 & 0.0072 &0.0271 &2.3658  &2.5851& 31.29 %31.11   
             \\
             & 2.236  &2.1758  &2.1702 & 0.0072 &0.0269 &2.3388 &2.5640& 31.27 %30.63
             \\ 
             &\multicolumn{7}{l}{N$^4$LO${^+}$+N$^2$LO ($\Lambda_N=500$ MeV)}  \\
             & 1.880   &  2.1996 & 2.1883&0.0072&0.0283&2.3571  &2.5653& 31.44 %31.25 
             \\
             & 2.236   &  2.1649 & 2.1590&0.0072&0.0281&2.3295  &2.5384& 31.47%30.68
             \\  
             &Exp. &&&&& & 2.517(30) 
             \cite{Tanihata:2013jwa} &31.99 \cite{Wang:2021xhn} \\
             &     &&&&& & 2.589(39) \cite{Nortershauser:2011zz} \\
             \hline
     $^{8}$Be  &\multicolumn{7}{l}{N$^4$LO${^+}$+N$^2$LO ($\Lambda_N=450$ MeV), $\omega_\text{HO}=18$ MeV, $N_\text{HO}=10$ }  \\
               & 1.880    & 2.1952 &2.1877&0.0074&0.0347 &2.3633 && 55.21 %51.36
               \\  
               &\multicolumn{7}{l}{N$^4$LO${^+}$+N$^2$LO ($\Lambda_N=500$ MeV), $\omega_\text{HO}=18$ MeV, $N_\text{HO}=10$ }  \\
               & 1.880    &2.1941 & 2.1821&0.0074&0.0372 &2.3605  && 55.18%%51.31 
               \\ 
               & Exp. &&&&&&& 56.50 \cite{Wang:2021xhn} \\
    \end{tabular}
    \end{ruledtabular}
\end{table*}

\begin{figure}[!htbp]
\centering
\includegraphics[width=\linewidth]{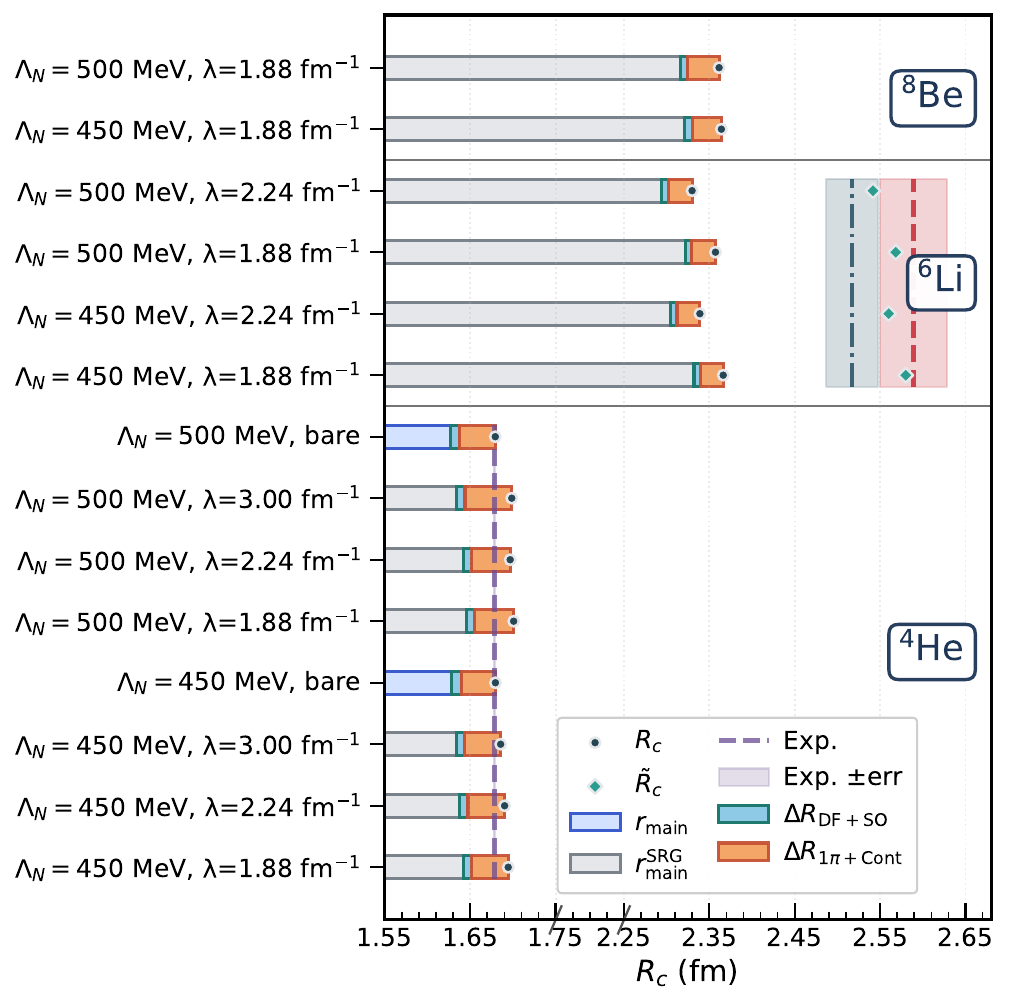}
\caption{Charge radii for $^{4}$He, $^6$Li and $^8$Be using the SMS interactions with $\Lambda_N=450$ and $500$ MeV corresponding to the result in Table~\ref{tab:rc}. The experimental data are indicated by the vertical lines with error bands. 
$r_{\text{main}}$ (FY) is obtained from the matter radius
$r_m$ , while
$r_{\text{main}}^{\text{SRG}}$ (J-NCSM) is obtained from the SRG-evolved matter radius
$r_m^{\text{SRG}}$; both include the proton and neutron charge radius contributions.
}
\label{fig:rc}
\end{figure}

The individual contributions to the charge radius from the point-proton radius $r_m$, DF and SO terms $\Delta R_\mathrm{DF+SO}$, and 2N contact and $1\pi$ operators $\Delta R_{\mathrm{Cont}+1\pi}$ are listed in Table~\ref{tab:rc} and shown in Fig.~\ref{fig:rc}. 
As mentioned before, the extracted $r_m$ does not 
account for the SRG effects 
when using the one-body density matrix corresponding to the SRG-evolved interaction. 
To evaluate this, we directly
calculate $\langle {\hat r}^2 \rangle $ using the SRG evolved two-body density at zero momentum
(see Ref. \cite{Sun:2025yfo} and Sec.~\ref{sec:S2} in t\cite{supp} for details). 
The resulting values are denoted by $r^\mathrm{SRG}_m$ and include
two-nucleon SRG-induced contributions to the radius operator. 
For $^{4}$He, the J-NCSM calculation yields converged radii and energies in good agreement with the FY approach with the bare interactions. 
We find that as the SRG flow parameter increases, the obtained radius decreases
and approaches the FY results, which 
by construction coincide with the most recent measurement
\cite{Krauth:2021foz}. 
The contributions from 2N operators with the SMS interactions for cutoffs $\Lambda_N=$450 and 500 MeV,  about 0.04~fm, 
are almost insensitive to the SRG transformation, visibly larger than the residual SRG dependence and are crucial for an accurate description of the charge radius. Notably, the importance of 2N contributions is consistent with preliminary results from a combined analysis of light nuclei ($A \le 4$), as reported in Ref.~\cite{Filin:2025Marciana}.

Our J-NCSM calculations for the $^6$Li charge radius do not yield a fully converged value; 
however, we find that the contributions from 2N charge operators exhibit stable convergence with increasing $N_\mathrm{HO}$ for different $\lambda$, see Fig.~\ref{fig:li6-r-2n} in the Ref. \cite{supp}. 
In particular, for a harmonic oscillator frequency of $\omega = 16$ or $18$~MeV, the 2N part of the charge radius, $\Delta R_{\mathrm{Cont}+1\pi}$, is almost converged with respect to $N_\mathrm{HO}$.
Therefore, we show the total charge radius and its decomposition into various operator contributions using this basis parameter set.
Inclusion of the 2N contributions increases the charge radius of ${}^6$Li by approximately 0.03~fm for both SMS interactions considered. Notably, using SRG-evolved two-body densities to evaluate the 2N operators can partially mitigate the dependence on the SRG flow parameter.
To further verify the influence of the 2N contribution for large mass number nuclei, we calculate the charge radius of 
$^8$Be,  although it is unbound with respect to the $\alpha$ emission and there is no available charge radius data.
For $^8$Be, the 2N contributions increase the charge radius by approximately 0.035 fm. 
It is remarkable that the size of the 2N contribution does neither scale with the number of 2N pairs nor with the number of $\alpha$ clusters in the nucleus. It seems that the contribution is, to a large extent, driven by short-distance currents getting contributions mainly from 4 nucleons in the $s$-shell. The $\alpha$-d ($\alpha$-$\alpha$) cluster structure of $^6$Li ($^8$Be) is not reflected in our explicit results of the 2N contributions.  
The importance of  the 2N contributions is clearly shown in Fig.~\ref{fig:rc}, where one finds that  $\Delta R_{\mathrm{Cont}+1\pi}$ is much larger than the DF and SO terms.
Recently, the role of two-body charge operators has also been investigated via variational Monte Carlo \cite{King:2024jiq,King:2025akz}, but using different chiral interactions.
The increase of $^6$Li and $^8$Be charge radii from two-body contributions found in our work is slightly larger than that in Ref. \cite{King:2025akz}.

%discussion of uncertainties
The uncertainties in our calculations 
primarily originate from three sources:
(a) the truncation of the J-NCSM model space and variation of the SRG flow parameter $\lambda$,
(b) the chiral order of the nuclear interactions and the corresponding charge currents and densities, and
(c) the LECs associated with the charge density operators.
For the $^6$Li charge radius, the truncation error of $\chi$EFT, estimated conservatively as a N$^2$LO contribution to $\tilde R_c$,
exceeds the spread arising from the variation of the cutoff $\Lambda$
(see Table \ref{tab:rc}), yet remains clearly smaller than the experimental error. The spread obtained from varying the SRG flow parameter $\lambda$
is comparable to this truncation estimate. The statistical uncertainty associated with the propagated uncertainties of the LECs  is negligible compared to 
all types of uncertainties, as demonstrated in~\cite{supp}. 
The basis space limitation in the radius calculations is treated following Ref.~\cite{Sun:2025yfo}, where a tail correction to the $r$-space two-body relative densities yields converged results. 
We apply this method to $^6$Li, obtaining charge radii denoted as $\tilde{R}_c$ in Table~\ref{tab:rc}, which are consistent with experimental measurements, see also Fig.~\ref{fig:rc}. 
For $^4$He, J-NCSM  provides converged radii and $R_c=\tilde R_c$.

{\it Summary}---Using high-order chiral SMS interactions and consistently regularized charge density operators, we study the charge form factors and radii of light isoscalar nuclei. 
We find that two-nucleon charge density contributions are much larger than those from DF and SO terms and are essential 
not only for describing the form factors at intermediate and large momentum transfers
but also for achieving an accurate determination of the charge radius. 
While the present study focuses on light isoscalar systems, 
the framework developed here can be extended to include consistently regularized 
isovector two-nucleon charge density operators (under development), 
and applied to heavier nuclei through alternative {\it ab initio} approaches. 
Our results suggest that two-nucleon charge operators could provide important contribution toward resolving the long-standing underestimation of nuclear charge radii in {\it ab initio} calculations. Their inclusion reduces the discrepancy between theory and experiment and is therefore essential for achieving quantitatively reliable predictions.

\begin{acknowledgments}
X.X.S.~appreciates  helpful discussion with Shan-Gui Zhou.
This work was supported in part by the European
Research Council (ERC) under the European Union's Horizon 2020 research
and innovation programme (grant agreement No. 101018170 and No. 885150)
as well as by the MKW NRW under the funding code NW21-024-A,
and by the CAS President's International Fellowship Initiative (PIFI) under Grant No.~2025PD0022.
 The numerical calculations were performed on JURECA
of the J\"ulich Supercomputing Centre, J\"ulich, Germany.
\end{acknowledgments}

\bibliographystyle{apsrev4-2}
\bibliography{ref.bib}
%\putbib
%\end{bibunit}

\clearpage

\setcounter{figure}{0}
\setcounter{table}{0}
\setcounter{equation}{0}
\renewcommand{\thefigure}{S\arabic{figure}}
\renewcommand{\thetable}{S\arabic{table}}
\renewcommand{\theequation}{S\arabic{equation}}

\section*{Supplemental materials of ``{\it Ab initio}  charge form factors and radii of light isoscalar nuclei: Role of the two-body charge density"}

\makeatletter
\renewcommand{\c@secnumdepth}{0}
\makeatother
\subsection{Calculation of charge form factors}
\label{sec:method}
The calculation of the charge form factor [cf. Eq. (\ref{eq:ffs})] is performed by
using the density matrix expansion method.
The one-body charge operator $\hat O_\mathrm{1b}$ including
the main contribution,
the  Darwin-Foldy (DF) and spin-orbit (SO) contributions
is calculated by \cite{Griesshammer:2020ufp} 
\begin{equation}
\begin{split}
\langle M' | {\hat O_\mathrm{1b}} | M \rangle & = 
\sum_{K=0}^{K_\text{max}=1}\sum_{\kappa=-K}^{K}\sum_{m_s,m'_s,m_\tau} 
{\rho_\text{1b}}^{M'M,m_\tau}_{K\kappa,m_sm'_s}(\vec k,\vec q) \\
&\times 
 {\hat O_\mathrm{1b}}(K\kappa; \vec q, \vec k; m_\tau;m_s m'_s),
\end{split}
\end{equation}
with the spherical expansion order $K$ and $\kappa$, and 
the spin-projection $m_s$ and isospin projection
$m_\tau$.

The contribution from the consistently regularized two-body charge density $\hat{O}_\text{2b}$
at next-to-next-to-next-to-leading order (N$^3$LO), consisting of 
contact (Cont) and one-pion-exchange ($1\pi$) terms, 
is 
\begin{equation}
\begin{split}
\langle M' | & {\hat O_\mathrm{2b}} | M \rangle  = 
\sum_{\alpha^{}_{12}\alpha'_{12}}\int \text{d} p_{12} \int \text{d}p'_{12}~
p_{12}^2 p_{12}^{\prime 2}\\
& \times
{\rho_{\text{2b}}}^{M'M}_{{\alpha'_{12}}\alpha^{}_{12}}(p'_{12}, p_{12};\vec q) {{\hat O}_\text{2b}}\phantom{x\! \! \! }  ^{M'M} 
_{{\alpha^{\prime}_{12} }{\alpha^{}_{12}}}
(p'_{12}, p_{12})
\end{split}
\end{equation}
where $\alpha\equiv (l_{12}s_{12})j_{12}t_{12}{m_t}_{12}$ labels the 
two-body partial-wave channels and the relative momentum is $p_{12}$ \cite{Griesshammer:2020ufp}.
Once the partial-wave decomposition is done, the resulting 
matrix elements of the charge operators are
used for all the nuclei.  
The nuclear many-body information 
is encoded in the one- and two-body density matrices,  constructed based on the J-NCSM calculations available online at 
\cite{densdb:2025nuc}.
The matrix elements of one- and two- charge density operator
are calculated by performing a numerical partial wave decomposition based on \cite{numpwd:2022}.

\subsection{Determination of isospin-1 channel LEC $M_3$}
\label{sec:S1}
There are three LECs, $M_1, M_2$, and $M_3$ in the  iso-scalar two-body contact charge density operators
\begin{equation}
\aligned
  \rho_\text{Cont}&(\bm k)  =
  2 e G_E^S(\bm{k}^2) \bigg[
    \ M_1 \frac{\bm{\sigma}_1\cdot \bm{\sigma}_2 + 3}{4}
    \frac{1- \bm{\tau}_1\cdot \bm{\tau}_2 }{4} 
     \bm k^2\\
    &+
    M_2
    \frac{1- \bm{\tau}_1\cdot \bm{\tau}_2 }{4}
    \left(
    (\bm k \cdot \sigma_1)(\bm k \cdot \sigma_2) - 
    \frac{1}{3} (\bm{\sigma}_1\cdot \bm{\sigma}_2)
     \bm k^2
    \right)
  \nonumber
  \\
    &
     +M_3\frac{1 - \bm{\sigma}_1\cdot \bm{\sigma}_2}{4}
        \left(
        \frac{\bm{\tau}_1\cdot \bm{\tau}_2 + 3}{4}
         \right) \bm k^2
  \bigg],
\endaligned
\end{equation}
where $\bm k$ is the photon momentum and $G_E^S(\bm{k}^2)$ is isoscalar nucleon charge form factor taken from \cite{Lin:2021xrc}. 
The details of the contact charge density operator
and its regularization can be found in Ref. \cite{Filin:2020tcs}.
LECs $M_1$, $M_2$, $M_3$ are related to LECs $A$, $B$, $C$ defined in \cite{Krebs:2020pii,Filin:2019eoe,Filin:2020tcs}
as
\begin{equation}
M_1 = A + B + C/3, \ M_2 =C, \ M_3=A - 3B - C.
\end{equation}
All LECs above have units of $F_\pi^{-2}\Lambda_b^{-3}$, where
$\Lambda_b=$~650~MeV refers to the breakdown scale of the chiral expansion and
$F_\pi = 92.4$ MeV is the pion decay constant.
The isospin-zero-to-isospin-zero channel LECs, 
$M_1$ and $M_2$, can be fitted to the deuteron charge and quadrupole form factors, and have already been
determined in Ref.~\cite{Filin:2020tcs}. 
In this work, to accurately determine the LEC $M_3$ in the isospin-1-to-isospin-1 channel
we use the 
high precision charge radius of $^4$He 
from muonic atom measurement 1.67824(83)~fm \cite{Krauth:2021foz}.

The charge radius of $^4$He is calculated via Eq.~\eqref{eq:rc}
using the one- and two-body density matrix from FY method with the
bare SMS interactions.
The matter radius contribution $r_m$  
is directly obtained using the two-body densities at zero momentum transfer.
The contributions from the DF, SO, and two-body one-pion-exchange and contact operators are 
extracted by using 
\begin{equation}
R_c^2 = -6\frac{\partial F(q^2)}{\partial q^2}\bigg|_{q^2=0}.
\label{eq:ff2rc}
\end{equation}  
In practice, the calculated FFs are fitted to the 
function $F(q^2) = 1 + aq^2 +bq^4+cq^6$ in the $q$ interval $(0, 0.5)$ $\mathrm{fm}^{-1}$ and 
the charge radius is extracted as $R_c = \sqrt{-6a}$.
We mention that the boost corrections are not included in our fitting procedure.

For the interaction SMS N$^{4}$LO$^{+}$ + N$^{2}$LO with $\Lambda_N=450$ MeV and $\Lambda_N=500$ MeV, 
the obtained values of the LEC $M_3$ read $M_3=(-2.44\pm0.13)$ $F_\pi^{-2}\Lambda_b^{-3}$ and $M_3=(-2.98\pm0.13)$ $F_\pi^{-2}\Lambda_b^{-3}$, respectively.

\subsection{Statistical uncertainties of charge radii propagated from LECs in 2N contact charge operator}

Since $M_3$ is fitted to the experimental $^4$He radius, which is known very  precisely, an inverse cross-check -- recomputing the $^4$He radius -- should reproduce an equally small uncertainty. In contrast, neglecting correlations between $M_1$ and $M_2$ for the interaction SMS 
N$^{4}$LO$^{+}$ + N$^{2}$LO with  cutoffs 450 MeV and 500 MeV
 yields  uncertainties of 0.0022 fm and 0.0019 fm, respectively.    These values are several times larger than the experimental precision.  Nevertheless, neglecting correlations is sufficient to provide a very conservative estimate for the uncertainty propagated to heavier nuclei, as the induced errors remain significantly smaller than other sources of uncertainty. 

Employing the three LECs discussed above, we examined the resulting uncertainties for the p-shell nuclei $^6$Li and $^{8}$Be. 
The J-NCSM calculations were performed with $\omega_\mathrm{HO}=18.0$ MeV and $N_\mathrm{HO}$=14, using the SMS interaction with $\Lambda=450$ MeV  evolved via SRG $\lambda =2.236$ $\mathrm{fm}^{-1}$, which 
reproduces the binding energy of $^6$Li. 
For $^6$Li, we obtain for the charge radius 
$
R_c= (2.3388\pm 0.0016_\text{stat.}) \, \mathrm{fm},
$
where the quoted uncertainty reflects the propagated uncertainty from the LECs. 
After tail correction, the charge radius is well reproduced.

A similar analysis was performed for  $^8$Be, using the J-NCSM calculations 
with $\omega_\mathrm{HO}=18.0$ MeV and $N_\mathrm{HO}$=10, and the SMS interaction with $\Lambda=450$ MeV SRG-evolved to 1.88 $\mathrm{fm}^{-1}$. 
In this case, the  uncertainty induced by the LECs 
is only about 0.0020 fm. 

In summary, the uncertainties from the LECs for $p$-shell nuclei are very small.
Therefore, in the main text, we show the results 
using the central values of the three LECs.

\subsection{Matter radius with the two-body relative density}
\label{sec:S2}
To assess the influence of the SRG transformation on the charge radius, we calculate the proton matter radius $r^\mathrm{SRG}_m$ of ${}^6$Li and ${}^8$Be
from the two-body relative densities in $r$-space. This is obtained by performing
the Fourier transformation on the two-body density matrix in momentum space, following the method introduced 
in Ref. \cite{Sun:2025aci}, where 
the SRG transformation is already accounted for by constructing the two-body transition density matrix 
with SRG evolved
nuclear many-body wave functions from J-NCSM calculations. 
The  matter radius is calculated using the  formulas introduced in Ref.~\cite{Sun:2025yfo}.
In this way, the SRG influence on charge radii
is considered at the two-body level and the induced three-body part is neglected.

\subsection{Charge form factors and radius of $^6$Li and $^8$Be from J-NCSM}
\label{Sec:S3}
\begin{figure}[tbp]
    \centering
    \includegraphics[width=0.9\linewidth]{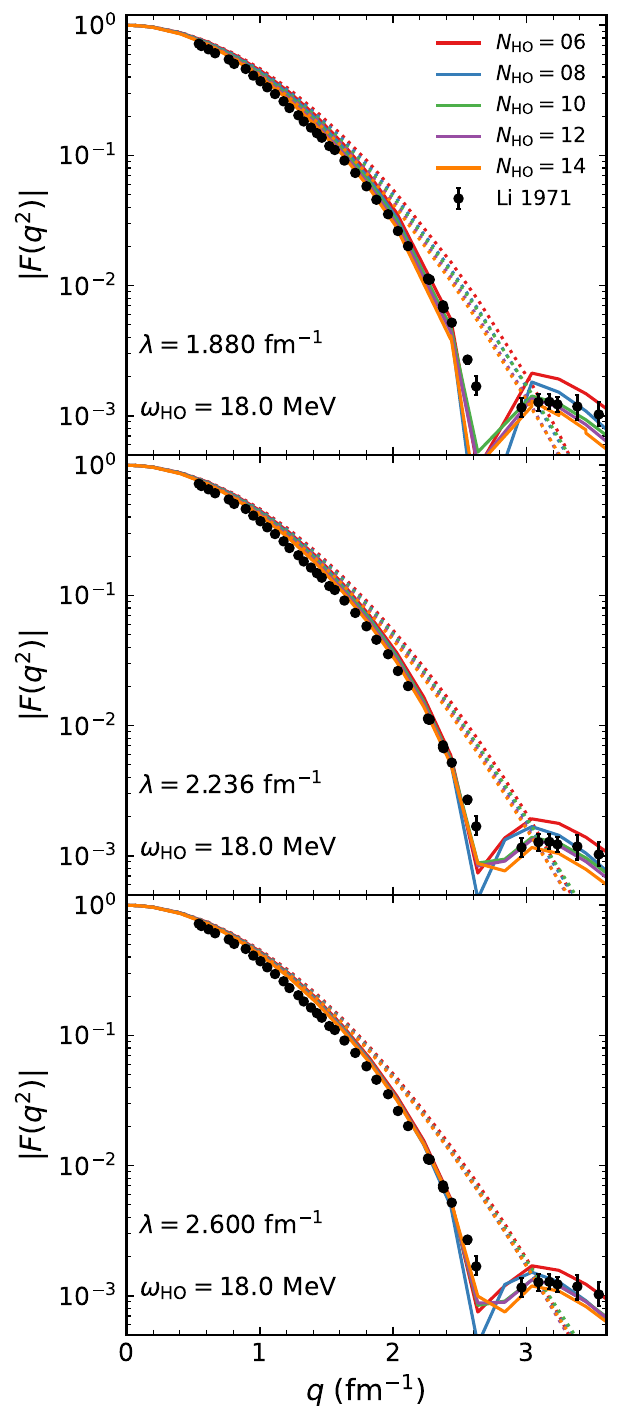}
    \caption{Charge form factor of $^6$Li from J-NCSM calculations with $\omega=18.0$ MeV and different $N_\mathrm{HO}$. The adopted interaction is SMS N$^{4}$LO$^{+}$ + N$^{2}$LO $\Lambda_N=$450 MeV evolved to flow parameters 1.88, 2.236 and 2.600 fm$^{-1}$. 
    The dotted lines present the results with one-body operators and solid lines for cases considering the 2N operators.
    Experimental data are taken from Ref. \cite{Li:1971tk}.
    }
    \label{fig:li6-ffs-1}
\end{figure}

\begin{figure}[tbp]
    \centering
    \includegraphics[width=0.9\linewidth]{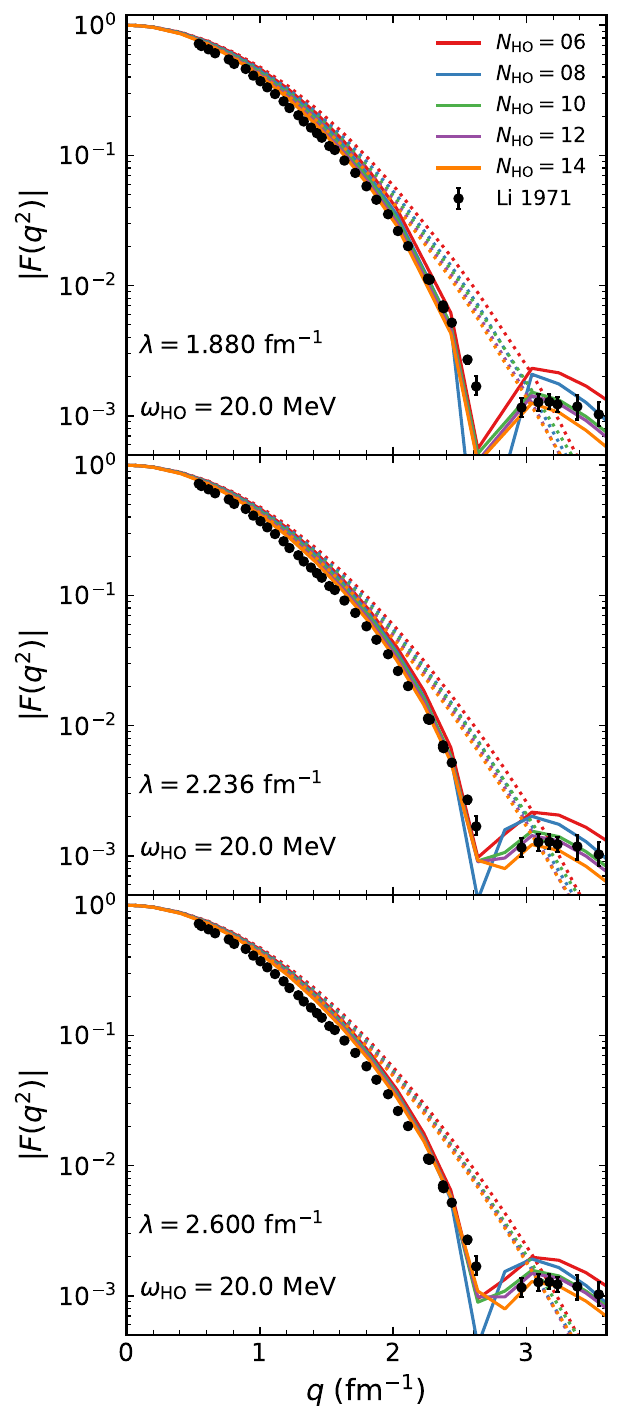}
    \caption{Charge form factor of $^6$Li from J-NCSM calculations with $\omega=20.0$ MeV and different $N_\mathrm{HO}$. The adopted interaction is SMS N$^{4}$LO$^{+}$ + N$^{2}$LO $\Lambda_N=$450 MeV evolved to flow parameter 1.88, 2.236, and 2.600 fm$^{-1}$.
    The dotted lines represent the results with one-body operators and solid lines for cases considering the 2N operators.
    Experimental data are taken from Ref. \cite{Li:1971tk}.
    }
    \label{fig:li6-ffs-2}
\end{figure}

In Figs.~\ref{fig:li6-ffs-1} and \ref{fig:li6-ffs-2}, we show the charge FFs of $^{6}$Li using J-NCSM with the chiral interaction SMS N$^{4}$LO$^{+}$ + N$^{2}$LO $\Lambda_N=$450 MeV evolved to flow parameters 1.88, 2.236 and 2.600 fm$^{-1}$. We show the results with
two $\omega_\mathrm{HO}$ values, 18.0 and 20.0 MeV, and $N_\mathrm{HO}=6,8,10,12$, and 14 and 
when $N_\mathrm{HO}=14$, the calculated binding energies are also close to the data. 
Especially when increasing SRG flow parameter $\lambda$,
the charge FFs in medium momentum transfers, 
$(2.0, 2.8)\  \mathrm{fm}^{-1}$, are closer to data.
Overall, with the increase of basis size $N_\mathrm{HO}$ and SRG flow parameter, the calculated FFs are well consistent with the measurements, and the two-body charge contribution plays an important role for high momentum transfers. Thereby, the effects from HO basis size and frequency are non-significant.

\begin{figure}[tbp]
    \centering
    \includegraphics[width=.9\linewidth]{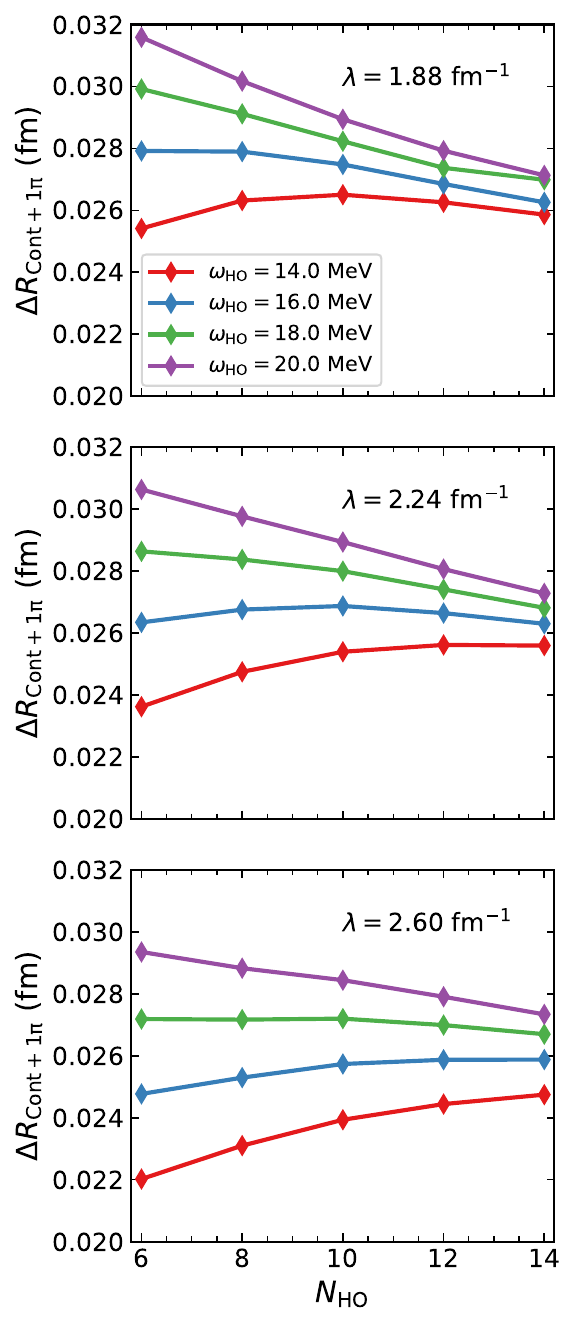}
    \caption{Contribution from two-nucleon contact and one-pion-exchange terms to the charge radii of $^6$Li from J-NCSM calculations with different HO parameters. The adopted interaction is SMS N$^{4}$LO$^{+}$ + N$^{2}$LO $\Lambda_N=$450 MeV evolved to flow parameter 1.88, 2.236 and 2.600 fm$^{-1}$.  }
    \label{fig:li6-r-2n}
\end{figure}

In Fig.~\ref{fig:li6-r-2n}, we show the contribution $\Delta R_{\mathrm{Cont}+1\pi}$ from the contact and one-pion-exchange operators to the charge radius of $^6$Li with different J-NCSM basis frequencies and model spaces  and for different SRG flow parameters. It can be seen that $\Delta R_{\mathrm{Cont}+1\pi}$ is converged with respect to  $N_\mathrm{HO}$ up to approximately 0.001~fm
especially when considering 
$\omega_\mathrm{HO}=$ 16 and 18 MeV, which are optimal for the convergence of the binding energy.

\begin{figure}[tbp]
    \centering
    \includegraphics[width=.9\linewidth]{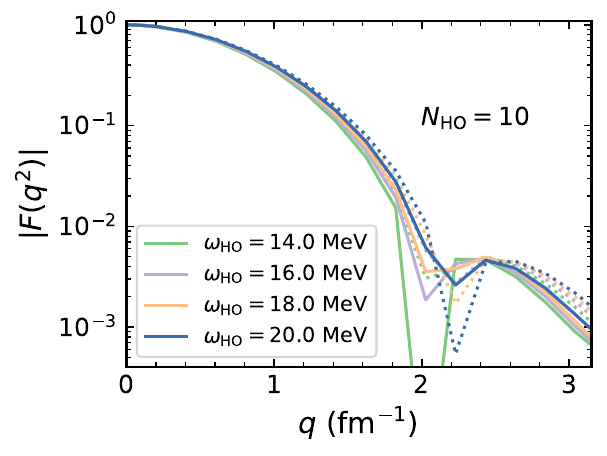}
    \caption{Charge form factor of $^8$Be from J-NCSM calculations with $N_\mathrm{HO}$=10
    and different $\omega$. The adopted interaction is SMS N$^{4}$LO$^{+}$ + N$^{2}$LO $\Lambda_N=$450 MeV evolved to flow parameter 1.88 fm$^{-1}$.
    The dashed lines indicate the results using 1N operators and the solid lines are those including 2N operator contributions. }
    \label{fig:be8-ffs}
\end{figure}

\begin{figure}[tbp]
    \centering
    \includegraphics[width=.9\linewidth]{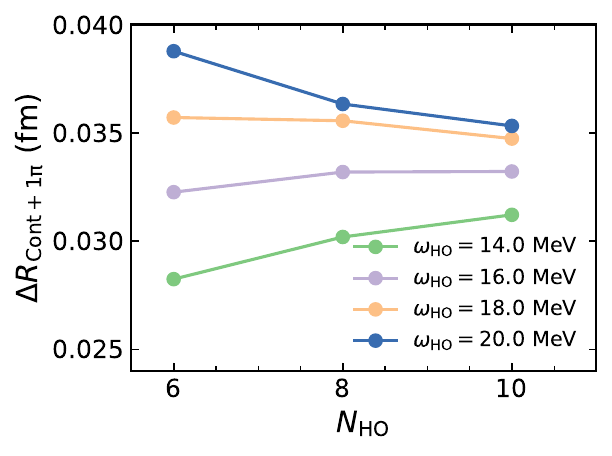}
    \caption{Contribution from two-nucleon contact and one-pion-exchange terms to the charge radii of $^8$Be from J-NCSM calculations with different HO parameters. The adopted interaction is SMS N$^{4}$LO$^{+}$ + N$^{2}$LO $\Lambda_N=$450 MeV evolved to flow parameter 1.88 fm$^{-1}$. }
    \label{fig:be8-r2n}
\end{figure}

For $^8$Be, similar calculations have also been  performed. 
In Fig.~\ref{fig:be8-ffs}, we show the charge FFs for four $\omega_\mathrm{HO}$ values using the interaction SMS N$^{4}$LO$^{+}$ + N$^{2}$LO $\Lambda_N=$450 MeV evolved to flow parameter 1.88 fm$^{-1}$. Due to the limited basis size in J-NCSM, the maximum HO model space is 10.
For the optimal HO frequencies 16.0 and 18.0 MeV, the differences in FFs are insignificant.
In Fig.~\ref{fig:be8-r2n}, we show the contribution of OPE and contact term on charge radius, $\Delta R_{\mathrm{Cont}+1\pi}$, as a function of $N_\mathrm{HO}$ for different $\omega_\mathrm{HO}$ and we find that for $\omega_\mathrm{HO}=16$ and $18$~MeV, $\Delta R_{\mathrm{Cont}+1\pi}$ are almost converged and the value is about $0.034 $~fm.

Calculations with J-NCSM using the interaction SMS N$^{4}$LO$^{+}$ + N$^{2}$LO for $\Lambda_N=500$ MeV yield conclusions for the effects of 2N charge operators on  $^6$Li and $^8$Be that are almost the same as for  $\Lambda=450$ MeV.

%%\putbib
%\bibliography{ref.bib}
%\end{bibunit}

\end{document}